\documentclass[10pt,a4paper]{article}
\usepackage{graphicx}
\usepackage{authblk}
\usepackage{amsfonts}

\usepackage[width=15cm, left=3cm]{geometry}

\usepackage{amssymb}

\usepackage{soul}
\usepackage{xcolor}

\newcommand{\beq}{\begin{equation}}
\newcommand{\eeq}{\end{equation}}
\newcommand{\beqa}{\begin{eqnarray}}
\newcommand{\eeqa}{\end{eqnarray}}

\def\ket#1{|\,#1\,\rangle}
\def\bra#1{\langle\, #1\,|}

\def\proj#1#2{\ket{#1}\bra{#2}}

\def\kpsi{\ket{\psi}}

\begin{document}

\title{The use of the output states generated by the  broadcasting of entanglement in quantum teleportation}
\author[]{ Iulia Ghiu$^{1,2}$\thanks{Corresponding author, email: iulia.ghiu@g.unibuc.ro}, C\u{a}t\u{a}lina C\^irneci$^1$, and George Alexandru Nemne\c{s}$^{1,2}$}
\affil[]{{$^1$University of Bucharest, Faculty of Physics, PO Box MG-11, R-077125, Magurele, Romania and \\
$^2$Research Institute of the University of Bucharest (ICUB), 90-92 Sos. Panduri, 5th District, Bucharest, Romania }}

\maketitle

\begin{abstract}
In this article, we find a theorem that gives a relation between the maximal fidelity of teleportation and the concurrence of the inseparable $X$ state used as a quantum channel in this process.  
Furthermore, we evaluate the concurrence of the output states generated by the local and nonlocal asymmetric broadcasting of entanglement and prove that the concurrence is greater in the case of nonlocal broadcasting. We analyze the possibility of using the output states obtained by the broadcasting of entanglement as quantum channels in quantum teleportation. We prove, with the help of the above-mentioned theorem, that all the inseparable states given by the local and nonlocal asymmetric broadcasting of entanglement are useful for quantum teleportation. Finally, we show that the maximal fidelity of teleportation is greater in the case when the second scenario is used, i.e., when the quantum channel is generated by the nonlocal asymmetric broadcasting of entanglement.
\end{abstract}


\section{Introduction}

In recent years, many quantum information processing protocols  have been used for developing applications that require quantum correlations. They represent new physical resources in quantum cryptography, quantum communication, quantum internet, and other applications.
Among quantum correlations, entanglement plays a fundamental role in quantum information theory \cite{Nielsenbook}, \cite{Horodecki-RMP-2009}, \cite{Horodecki-PRL-2002}, \cite{Jin-2015}, \cite{Ghiu-2001}, \cite{Sen}, \cite{Ghiu-rus-2014}, \cite{Wang-2022}, \cite{Ghiu-2013}, \cite{Nosrati}, \cite{Nosrati-q-inf-2020}, \cite{Ghiu-2012}, \cite{Paris}.

By using a maximally entangled state of a two-qubit system shared between two distant observers, Alice and Bob, Bennett {\it et al.} proved that quantum teleportation can be achieved with unity fidelity \cite{Bennett-1993}. Further, this protocol was extended to the cases when many receivers were involved (one-to-many teleportation) \cite{Murao-2000} and, more generally, many senders and many receivers participated (many-to-many teleportation) \cite{Ghiu-pra-2003}.

Another scenario that was investigated is when a mixed two-qubit state is employed as a quantum channel in the standard teleportation protocol. The purpose was to find the mixed two-qubit states  that generate a fidelity of teleportation that is higher than the 2/3 obtained in the case when a classical channel was used. Horodecki {\it et al.} presented a theorem for finding the necessary and sufficient condition for a two-qubit state to be useful for standard teleportation \cite{Horodecki-PLA-1996}. It was demonstrated  that not all the inseparable mixed two-qubit states can be used as quantum channels in  teleportation.
With the help of this theorem,  the case where the Werner state was used as a quantum channel in teleportation was examined, as well as different other two-qubit mixed states \cite{Adhikari-2010}.
The teleportation of a qubit's state was recently investigated using a non-maximally entangled two-qudit ($d$-level system) state as a quantum channel (with $d\ge 3$), and it was demonstrated that the teleportation is realized with unity fidelity \cite{Chen-2022}.

Local symmetric broadcasting of entanglement was introduced in Ref. \cite{Buzek-1997} and was realized by applying the local $1\to 2$ optimal universal symmetric cloning machine \cite{Bruss-1998} on the initial pure entangled state by each observer. The $1\to 2$ symmetric cloning machine generates two identical output states.
A different scenario was considered in Ref. \cite{Buzek-prl-1998}, where a nonlocal optimal universal symmetric cloning machine was used on the initial pure entangled state in order to copy the entanglement. It was demonstrated that the range of the parameter $\alpha $ in the case of nonlocal broadcasting of entanglement is larger than the range of $\alpha $ when local symmetric broadcasting of entanglement is performed. A nice generalization of  the broadcasting of entanglement was considered by using orthogonal and
non-orthogonal state-dependent cloning machines \cite{Chatterjee-qip-2020}.

Cerf proposed a generalization of the $1\to 2$ optimal universal cloning machine by considering two different output states, whose
Bloch vector is in the same direction as the initial state, and creating the second clone with maximal fidelity for a given fidelity of the first one. This was called the $1\to 2$ optimal universal asymmetric cloning machine \cite{Cerf-prl-2000}, \cite{Cerf-jmo-2000}. One of us introduced the local and nonlocal asymmetric broadcasting of the entangled two-qubit pure state by employing the asymmetric cloner \cite{Ghiu-pra-2003}. This concept was recently extended to the study of asymmetric broadcasting of an arbitrary entangled two-qubit mixed state \cite{Chatterjee-pra-2019}. We want to emphasize that for evaluating the inseparability in the process of broadcasting of entanglement in the above references, only the Peres-Horodecki criterion \cite{Peres}, \cite{Horodecki-PLA-2-1996} was used and that no measure of entanglement was computed.

The aim of the present article is twofold. First, we want to find a relation between the maximal fidelity of teleportation and the concurrence of the quantum channel needed in this process, for a special class of two-qubit mixed states. Second, we investigate the possibility of using the output states generated by the local and nonlocal asymmetric broadcasting of the entanglement as quantum channels for quantum teleportation. The paper is organized as follows: In Section 2, we prove a theorem that presents the dependence of the maximal fidelity of teleportation on the concurrence of a special class of inseparable X states used as quantum channels in this process.
Section 3 is dedicated to evaluating the concurrence of the output states generated by the local broadcasting of entanglement. In addition, we obtain the conditions for both output states to be inseparable. In Section 4, we compute the concurrence of the two given states via nonlocal broadcasting of entanglement and determine the conditions needed for these states to be inseparable. Further, we prove in Section 5 that the concurrence of the inseparable states obtained by using nonlocal broadcasting is greater than the concurrence of the states generated via local broadcasting. In Section 6, we show that all the inseparable states generated in the process of local and nonlocal broadcasting of entanglement can be used as quantum channels in teleportation. We prove in Section 7 that the fidelity of teleportation is greater in the case of the inseparable states obtained in nonlocal broadcasting than in the case of local broadcasting of entanglement.
Our conclusions are outlined in Section 8.

\section{Fidelity of teleportation for X states and its relation to concurrence}

Consider that the state of the qubit to be teleported from Alice to Bob is a pure one described by the projector $P_\phi$.
The channel used in quantum teleportation, which is shared by Alice and Bob, is found in a two-qubit mixed state described by the density operator $\rho $. This means that the three-particle system is initially found in the state $P_\phi \otimes \rho $. Alice performs a measurement on her particles in the Bell basis, whose elements are described by the projectors $P_k$, in the standard quantum teleportation. The probability that Alice will obtain the outcome $k$ is given by \cite{Horodecki-PLA-1996}:
\beq
p_k=\mbox{Tr}\left[ (P_k\otimes I)(P_\phi \otimes \rho )  \right].
\eeq

The best strategy determined by the unitary operators $U_k$ must then be applied by Bob in order to achieve the highest fidelity. The state of Bob at the end of teleportation is denoted by $\rho_k$ and is given by \cite{Horodecki-PLA-1996}:
\beq
\rho_k=\frac{1}{p_k}\, \mbox{Tr}_{1,2}\left[ (P_k\otimes U_k) (P_\phi \otimes \rho ) (P_k\otimes U_k^\dagger) \right].
\eeq

The fidelity of teleportation is given by \cite{Horodecki-PRA-1996}:
\beq
F(\rho )=\int \sum_{k=1}^4 p_k\, \mbox{Tr} \left( \rho_k P_\phi \right)\, d\, M(\phi ).
\label{gen-fid}
\eeq
In the above expression, the integral is performed over all the pure states $\ket{\phi}$ of the Bloch sphere, which are characterized by a uniform distribution $M$.

The density operator of a two-qubit system can be written as follows \cite{Fano}, \cite{Horodecki-PRA-1996}:
\beq
\rho = \frac{1}{4}\, \left( I\otimes I + {\bf r}\cdot {\bf \sigma} \otimes I+ I \otimes {\bf s}\cdot {\bf \sigma} +\sum_{m,n=1}^3T_{mn}\, \sigma_m \otimes \sigma_n \right),
\label{op-gen}
\eeq
where $\sigma_j$, with $j$ = 1, 2, 3 represent the Pauli operators. Eq. (\ref{op-gen}) is called the Fano parametrization.
The vectors ${\bf r}$ and ${\bf s}$ are real, their expressions being given by $r_j=\mbox{Tr}(\rho \, \sigma_j\otimes I)$ and $s_j=\mbox{Tr}(\rho \, I\otimes \sigma_j)$. The $3 \times 3$  matrix $T$, having its elements defined by $T_{mn}$, is a real matrix, where $T_{mn}=\mbox{Tr}(\rho \, \sigma_m\otimes \sigma_n)$ with $m,n$ = 1, 2, 3.
If one defines the function \cite{Horodecki-PLA-1996}
\beq
N(\rho ) = \mbox{Tr}\sqrt{T^T\, T},
\label{N-ro}
\eeq
then, as presented in Ref. \cite{Horodecki-PLA-1996}, we have the following Theorem:

An arbitrary mixed two-qubit state can be used as a quantum channel in the standard teleportation iff
\beq
N(\rho )>1.
\label{cond-telep}
\eeq
In addition, by using Eq. (\ref{gen-fid}), it was proved that the maximal fidelity of teleportation, when the two-qubit state $\rho $ is used as a quantum channel, has the following expression \cite{Horodecki-PLA-1996}:
\beq
F_{\mbox{max}}(\rho )=\frac{1}{2}\left[ 1+\frac{1}{3}\, N(\rho) \right].
\label{fid-telep}
\eeq
This means that there are inseparable states characterized by $N(\rho )\le 1$, which cannot be used as a quantum channel in the standard quantum teleportation.

Let us investigate the particular case when the quantum channel $\rho $ used in teleportation is found in the $X$ state \cite{Rau}, \cite{Ghiu-entropy}, \cite{Ghiu-results}:
\beq
\rho_{\mbox x}=\left(
\begin{array}{cccc}
 \rho_{11} & 0 & 0 & \rho_{14} \\
 0 & \rho_{22} & \rho_{23} & 0 \\
 0 & \rho_{32} & \rho_{33} & 0 \\
 \rho_{41} & 0 & 0 & \rho_{44} \\
\end{array}
\right).
\label{x-st}
\eeq
The $X$ states are described by non-zero elements only along the diagonal and the anti-diagonal. The above $\rho_{jj}$ are real parameters ($j$ = 1, 2, 3, 4) and the off-diagonal terms are complex parameters. For an $X$ state denoted by $\rho_X$ the Fano parametrization reads \cite{Ghiu-entropy}:
\beqa
{\bf r_{\mbox x}}:&& 0, 0, r; \nonumber \\
{\bf s_{\mbox x}}:&& 0, 0, s; \label{fano-x} \\
T_{\mbox x}&=&\left( \begin{array}{ccc}
T_{11}&T_{12}&0\\
T_{21}&T_{22}&0\\
0&0&T_{33}
\end{array} \right). \nonumber
\eeqa

The elements of the matrix $T$ have the following expression \cite{Rau}, \cite{Ghiu-entropy}:
\beqa
T_{11}&=&2\, {\rm Re}[\rho_{23}+\rho_{14}], \nonumber\\
T_{22}&=&2\, {\rm Re}[\rho_{23}-\rho_{14}], \nonumber\\
T_{33}&=&\rho_{11}-\rho_{22}-\rho_{33}+ \rho_{44}, \label{Fano-X} \\
T_{12}&=&2\, {\rm Im}[\rho_{23}-\rho_{14}],\nonumber\\
T_{21}&=&-2\, {\rm Im}[\rho_{23}+\rho_{14}].\nonumber
\eeqa

Suppose, in addition, that the $X$ state used as a quantum channel in teleportation is characterized by $\rho_{23} =0$.
Then, according to Eqs. (\ref{Fano-X}), one obtains:
\beqa
T_{12}&=&T_{21};\nonumber \\
T_{11}&=&-T_{22}.
\eeqa

By using definition (\ref{N-ro}), we obtain the expression of the function $N(\rho)$ in the case of an $X$ state with $\rho_{23} =0$ as follows:
\beqa
N(\rho)& =& 2\, \sqrt{T_{11}^2+T_{12}^2}+ |T_{33}|\nonumber \\
&=&4\, |\rho_{14}|+|1-2\, \rho_{22}-2\, \rho_{33}|.
\label{N-ro-particular}
\eeqa

In our paper, we employ the so-called concurrence in order to measure the entanglement of the two qubits.  Wootters introduced the definition of the concurrence in the case of a pure two-qubit state
$\ket{\phi}= a\, \ket{00} + b\, \ket{01} + c\, \ket{10} + d\, \ket{11}$ as follows \cite{Wootters-1998}, \cite{Wootters-2001}:
\beq
C(\ket{\phi})=2\, |a\, d-b\, c|.
\label{conc-st-pura}
\eeq

In addition, in the case of a mixed two-qubit state described by the density operator $\rho $, one has to introduce the spin-flipped state as being: $\rho '= (\sigma_2\otimes \sigma_2)\, \rho^*\, (\sigma_2\otimes \sigma_2)$, where $\rho^*$ is the complex conjugate of $\rho $.
One can prove that $\rho \, \rho '$ is a non-Hermitian matrix \cite{Wootters-1998}, \cite{Wootters-2001}, being characterized by four eigenvalues, which are real and non-negative \cite{Fan-2019}. If $\nu _1$, $\nu _2$, $\nu _3$, and $\nu _4$ are the eigenvalues of the matrix $\rho \, \rho '$ written in decreasing order, then the concurrence has the expression \cite{Wootters-2001}:
\beq
C(\rho)=\max\{\sqrt{\nu_1}-\sqrt{\nu_2}-\sqrt{\nu_3}-\sqrt{\nu_4},0 \}.
\eeq

For the special two-qubit states, which are described by $X$ states, the concurrence has the following formula \cite{Eberly-2007}:
\beq
C(\rho_{\mbox x})=2\, \max \left\{ 0, |\rho_{23}|-\sqrt{\rho_{11}\, \rho_{44}}, \, |\rho_{14}|-\sqrt{\rho_{22}\, \rho_{33}}\right\}.
\label{conc-X}
\eeq

{\bf Theorem.} {\it Any inseparable $X$ state characterized by $\rho_{23}=0$ and $\rho_{22}=\rho_{33}< \frac{1}{4}$ is useful for the standard teleportation. The relation between the maximal fidelity of teleportation and the concurrence for such states is:}
\beq
F_{\mbox{max}}(\rho_{\mbox x})= \frac{2}{3}+\frac{1}{3}\, C(\rho_{\mbox x}).
\label{leg-fid-concurr}
\eeq

{\bf Proof.} According to Eq. (\ref{conc-X}), the concurrence of an $X$ state with $\rho_{23}=0$ and $\rho_{22}=\rho_{33}$ has the expression:
\beq
C(\rho_{\mbox x})=2\, \max \left\{ 0, \, |\rho_{14}|-\rho_{22}\right\}.
\label{conc-X-particular}
\eeq
This leads to the condition that the $X$ state must be inseparable:
\beq
|\rho_{14}|>\rho_{22}.
\label{cond-insep-X}
\eeq

By using the condition $\rho_{22}=\rho_{33}$ and applying Eq. (\ref{N-ro-particular}), we determine the expression of $N$:
\beq
N(\rho_{\mbox x})=4\, |\rho_{14}|+|1-4\, \rho_{22}|=1+ 4\, (|\rho_{14}|- \rho_{22}),
\eeq
where above we used $\rho_{22}< \frac{1}{4}$. The $X$ states are inseparable if the inequality (\ref{cond-insep-X}) is satisfied, which leads to
\beq
N(\rho_{\mbox x})>1,
\eeq
i.e. the state is useful for the standard teleportation according to Eq.  (\ref{cond-telep}). With the help of Eq. (\ref{conc-X-particular}), we obtain the expression of the function $N$ in terms of the concurrence:
\beq
N(\rho_{\mbox x})=1 + 2\, C(\rho_{\mbox x}).
\eeq
Having the formula of $N$, by using Eq. (\ref{fid-telep}) we arrive at the final relation between the maximal fidelity of teleportation and the concurrence:
\beq
F_{\mbox{max}}(\rho_{\mbox x})= \frac{2}{3}+\frac{1}{3}\, C(\rho_{\mbox x}),
\eeq
which ends the proof of the Theorem.

\section{Broadcasting of entanglement using local optimal universal asymmetric cloning machines}
\subsection{The output states generated by the local asymmetric broadcasting of entanglement}

Suppose that the initial entangled two-qubit system shared by Alice and Bob is
\beq
\kpsi_{a_1b_1}=\alpha \ket{00}+\beta \ket{11},
\label{entini}
\eeq
where $\alpha$ and  $\beta$ are complex parameters satisfying $|\alpha|^2+|\beta|^2=1$.
In addition, Alice holds two ancillary systems found in the states $\ket{0}_{a_2}$ and $\ket{0}_{a_3}$, while Bob has two other particles found in the states $\ket{0}_{b_2}$ and $\ket{0}_{b_3}$.

The optimal universal asymmetric cloning machine acting on qubits is described by the following unitary operator \cite{Ghiu-pra-2003}:
\beqa
U(p)\ket{0}\ket{00}=\frac{1}{\sqrt{1+p^2+q^2}}(\ket{000}+p\ket{011}+q\ket{101}),\nonumber\\
U(p)\ket{1}\ket{00}=\frac{1}{\sqrt{1+p^2+q^2}}(\ket{111}+p\ket{100}+q\ket{010}),
\label{Pauli}
\eeqa
with $p+q=1$. The particular case $p=1/2$ leads to the optimal universal symmetric cloning machine.

Suppose that each Alice and Bob apply the optimal universal asymmetric cloning machine described by the same parameter $p$ on the total initial state $\kpsi_{a_1b_1}\ket{00}_{a_2a_3}\ket{00}_{b_2b_3}$:
\beqa
\ket{\phi}&=&U(p)\otimes U(p)\kpsi_{a_1b_1}\ket{00}_{a_2a_3}\ket{00}_{b_2b_3}\nonumber\\
&=&\frac{1}{1+p^2+q^2}\{\ket{00}_{a_3b_3}[\alpha \ket{00}_{a_1a_2}\ket{00}_{b_1b_2}
+\beta p^2\ket{10}_{a_1a_2}\ket{10}_{b_1b_2}\nonumber\\
&&+\beta pq\ket{10}_{a_1a_2}\ket{01}_{b_1b_2}
+\beta pq\ket{01}_{a_1a_2}\ket{10}_{b_1b_2}+\beta q^2\ket{01}_{a_1a_2}\ket{01}_{b_1b_2}]\nonumber\\
&&+\ket{01}_{a_3b_3}[\alpha p\ket{00}_{a_1a_2}\ket{01}_{b_1b_2}+\alpha q\ket{00}_{a_1a_2}\ket{10}_{b_1b_2}+\beta p\ket{10}_{a_1a_2}\ket{11}_{b_1b_2}\nonumber\\
&&+\beta q\ket{01}_{a_1a_2}\ket{11}_{b_1b_2}]
+\ket{10}_{a_3b_3}[\alpha q\ket{10}_{a_1a_2}\ket{00}_{b_1b_2}+\alpha p\ket{01}_{a_1a_2}\ket{00}_{b_1b_2}\nonumber\\
&&+\beta p\ket{11}_{a_1a_2}\ket{10}_{b_1b_2}+\beta q\ket{11}_{a_1a_2}\ket{01}_{b_1b_2}]
+\ket{11}_{a_3b_3}[\alpha p^2\ket{01}_{a_1a_2}\ket{01}_{b_1b_2}\nonumber\\
&&+\alpha pq\ket{01}_{a_1a_2}\ket{10}_{b_1b_2}
+\alpha pq\ket{10}_{a_1a_2}\ket{01}_{b_1b_2}+\alpha
q^2\ket{10}_{a_1a_2}\ket{10}_{b_1b_2}\nonumber\\
&&+\beta \ket{11}_{a_1a_2}\ket{11}_{b_1b_2}]\}.
\eeqa

The two states shared by Alice and Bob are given by the reduced density operators:
\beqa
&&\rho^{a_1b_1}=\mbox{Tr}_{a_2b_2a_3b_3}\proj{\phi}{\phi}\label{st-a1b1-loc} \\
&&=\frac{1}{(1+p^2+q^2)^2}\left( \begin{array}{cccc}
|\alpha|^2(1+p^2)^2+|\beta|^2q^4&0&0&4\alpha \beta^*p^2\\
0&q^2(1+p^2)&0&0\\
0&0&q^2(1+p^2)&0\\
4\alpha^* \beta p^2&0&0&|\beta|^2(1+p^2)^2+|\alpha|^2q^4
\end{array} \right);\nonumber
\eeqa

\beqa
&&\rho^{a_2b_2}=\mbox{Tr}_{a_1b_1a_3b_3}\proj{\phi}{\phi}\label{st-a2b2-loc} \\
&&=\frac{1}{(1+p^2+q^2)^2}\left( \begin{array}{cccc}
|\alpha|^2(1+q^2)^2+|\beta|^2p^4&0&0&4\alpha \beta^*q^2\\
0&p^2(1+q^2)&0&0\\
0&0&p^2(1+q^2)&0\\
4\alpha^* \beta q^2&0&0&|\beta|^2(1+q^2)^2+|\alpha|^2p^4
\end{array} \right).\nonumber
\eeqa
According to Eq. (\ref{x-st}), the two output states $\rho^{a_1b_1}$ and $\rho^{a_2b_2}$ are $X$ states.

The two-qubit states, which belong only to Alice and only to Bob, respectively, are obtained as the reduced density operators
\beqa
\rho^{a_1a_2}=\mbox{Tr}_{b_1b_2a_3b_3}\proj{\phi}{\phi};\nonumber \\
\rho^{b_1b_2}=\mbox{Tr}_{a_1a_2a_3b_3}\proj{\phi}{\phi},
\eeqa
their formulae being
\beqa
&&\rho^{a_1a_2}=\rho^{b_1b_2}=\frac{1}{(1+p^2+q^2)^2}\nonumber \\
&&\times \left( \begin{array}{cccc}
|\alpha|^2(1+p^2+q^2)&0&0&0\\
0&p^2q^2+|\beta|^2q^4+|\beta|^2q^2&pq+p^3q+pq^3&0\\
&+|\alpha|^2p^4+|\alpha|^2p^2 &&\\
0&pq+p^3q+pq^3&p^2q^2+|\beta|^2p^4+|\beta|^2p^2&0\\
&&+|\alpha|^2q^4+|\alpha|^2q^2 &\\
0&0&0&|\beta|^2(1+p^2+q^2)
\end{array} \right).\nonumber
\eeqa
We remark also that the two local states $\rho^{a_1a_2}$ and $\rho^{b_1b_2}$ are $X$ states.

\subsection{The concurrence of the output states generated by the local asymmetric broadcasting of entanglement}

The concurrence of the local states $\rho^{a_1a_2}$ and $\rho^{b_1b_2}$ is evaluated with the help of formula (\ref{conc-X}):
\beq
C(\rho^{a_1a_2})=C(\rho^{b_1b_2})=2\max \bigg\{ 0, \frac{pq-|\alpha |\, |\beta |}{1+p^2+q^2} \bigg\}.
\eeq
The condition for the two local states to be separable is given by $|\alpha |\, |\beta |\ge p\, q$ or equivalently
\beq
\frac{1}{2}\left[ 1-\sqrt{1-4p^2(1-p)^2}\right]\le |\alpha |^2 \le \frac{1}{2}\left[ 1+\sqrt{1-4p^2(1-p)^2}\right].
\label{cond-sep}
\eeq

We evaluate now the concurrence of the states shared by Alice and Bob by using Eq. (\ref{conc-X}):
\beqa
C(\rho^{a_1b_1})&=&2\max \bigg\{ 0, \frac{4p^2|\alpha |\, |\beta |-q^2(1+p^2)}{(1+p^2+q^2)^2}  \bigg\} \label{conc-loc-a1b1}\\
C(\rho^{a_2b_2})&=&2\max \bigg\{ 0, \frac{4q^2|\alpha | \, |\beta |-p^2(1+q^2)}{(1+p^2+q^2)^2} \bigg\}.\label{conc-loc-a2b2}
\eeqa

\begin{figure}[h!]
\centering
\includegraphics[width=6cm]{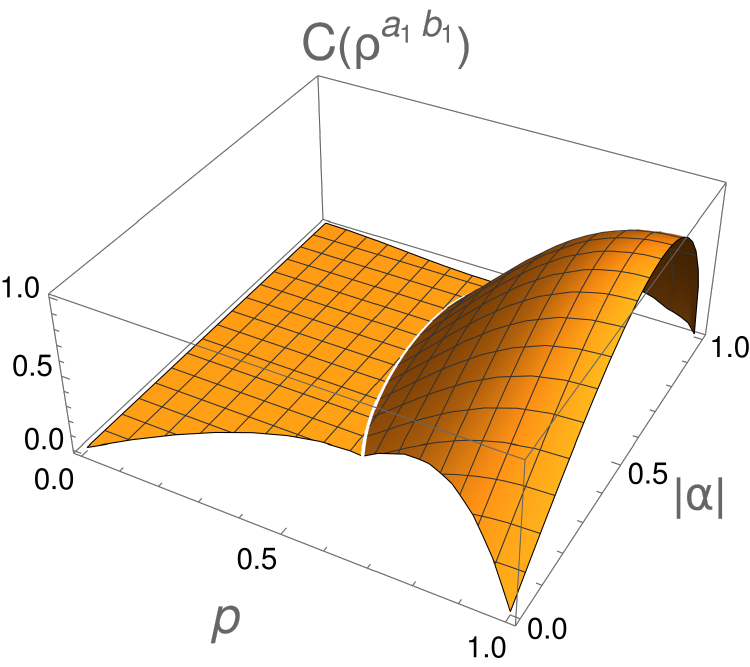}
\includegraphics[width=6cm]{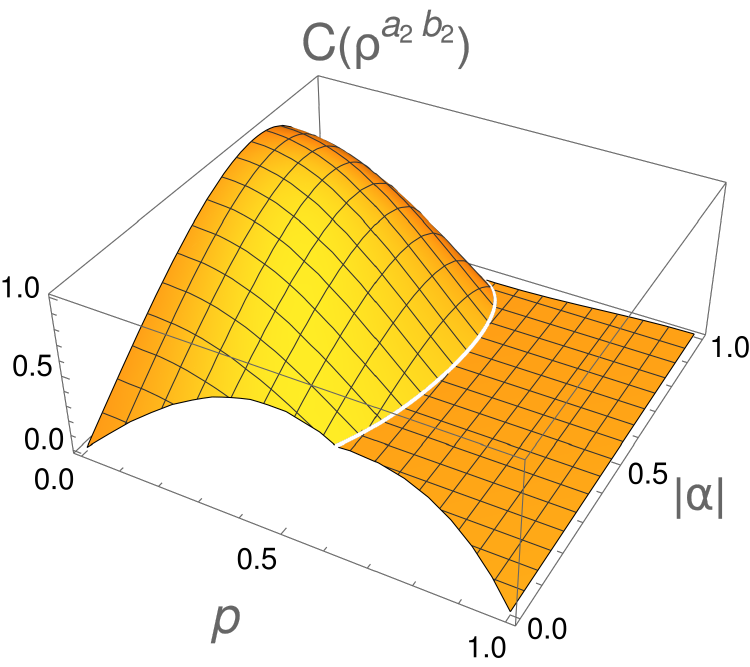}
\caption{Concurrence of $ \rho^{a_1b_1}$ (left) and $ \rho^{a_2b_2}$ (right) using local asymmetric cloning machines.}
\label{fig-conc-loc-a1-b1}
\end{figure}

In Fig. \ref{fig-conc-loc-a1-b1} we plot the concurrence of the states $\rho^{a_1b_1}$ and $\rho^{a_2b_2}$, respectively, in terms of $|\alpha |$ and $p$ by imposing the condition (\ref{cond-sep}).

\begin{figure}[h!]
\centering
\includegraphics[width=6cm]{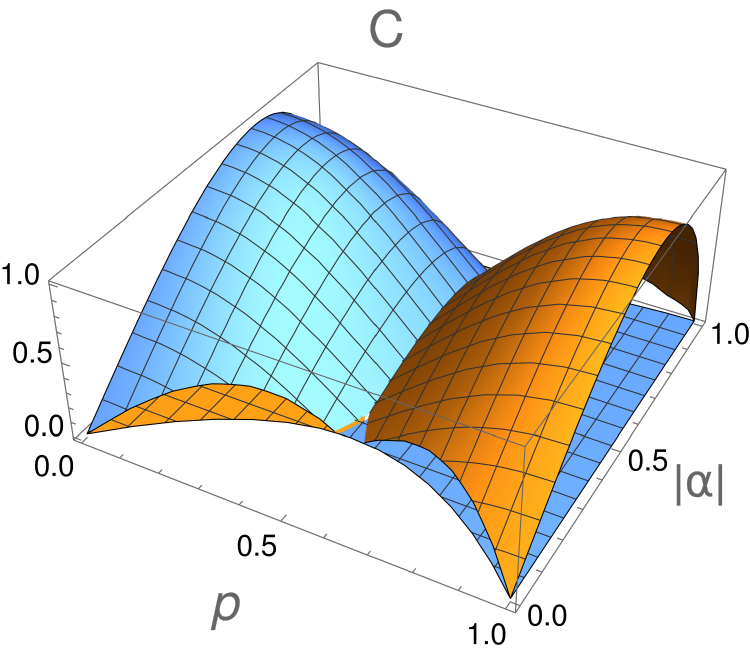}
\caption{A comparison between the concurrence of $ \rho^{a_1b_1}$ and the concurrence of $ \rho^{a_2b_2}$ when local asymmetric cloning machines are employed.}
\label{fig-conc-loc-a1b1-a2b2}
\end{figure}

Let us find the condition for the maximum of the concurrence of these two states. Therefore, we evaluate the partial derivative of the concurrence with respect to $|\alpha|$:
\beqa
\frac{\partial C(\rho^{{a_1}{b_1}})}{\partial |\alpha|}&=&\frac{2\left(1-2 |\alpha|^2\right) p^2}{\sqrt{1-|\alpha|^2}\left(1-p+p^2\right)^2}; \nonumber \\
\frac{\partial C(\rho^{{a_2}{b_2}})}{\partial |\alpha|}&=&\frac{2\left(1-2 |\alpha|^2\right) q^2}{\sqrt{1-|\alpha|^2}\left(1-p+p^2\right)^2}.
\eeqa
In both cases we have that the maximum of the concurrence is obtained when $|\alpha|=\frac{1}{\sqrt{2}}$, i.e. when the initial state (\ref{entini}) is the maximally entangled state.

We are interested in finding some conditions for $|\alpha |$ and $p$ such that the concurrence of both states $\rho^{a_1b_1}$ and $\rho^{a_2b_2}$ are simultaneously positive. This will be found as the intersection of the positive surfaces represented by the two concurrences, as shown in Fig. \ref{fig-conc-loc-a1b1-a2b2}.

From Eq. (\ref{conc-loc-a1b1}), it follows that the condition
 $C(\rho^{a_1b_1})> 0$ is equivalent to:
\beq
|\alpha |^4-|\alpha |^2+\frac{1}{16\, p^4}\, (1-p)^4(1+p^2)^2< 0.
\label{ec-p-1}
\eeq
Inequality (\ref{ec-p-1}) is satisfied only if $p\in (p_1, 1]$, where
\beq
p_1=\frac{1}{2}-\frac{3^{1/4}}{\sqrt 2}+\frac{\sqrt 3}{2} \simeq 0.435
\label{expr-p1}
\eeq
is a root of the equation $p^4-2\, p^3-2\, p+1=0$. If we denote
\beq
f _{\pm }=\sqrt{\frac{1}{2}\left[ 1\pm \sqrt{1-\frac{1}{4p^4}(1-p)^4(1+p^2)^2} \right] } ,
\eeq
then inequality (\ref{ec-p-1}) is verified for
\beq
|\alpha |\in (f_-,f_+).
\label{alfa-local-1}
\eeq

By using Eq. (\ref{conc-loc-a2b2}), one finds that the condition $C(\rho^{a_2b_2})> 0$ reads:
\beq
|\alpha |^4-|\alpha |^2+\frac{1}{16\, q^4}\, p^4(1+q^2)^2< 0,
\label{ec-p-2}
\eeq
which requires $p\in [0, p_2)$, where
\beq
p_2=\frac{1}{2}+\frac{3^{1/4}}{\sqrt 2}-\frac{\sqrt 3}{2} \simeq 0.565
\label{expr-p2}
\eeq
 is a root of the equation $p^4-2\, p^3+4\, p-2=0$. With the notation
\beq
g _{\pm }=\sqrt{\frac{1}{2}\left[ 1\pm \sqrt{1-\frac{1}{4q^4}p^4(1+q^2)^2} \right]},
\eeq
condition (\ref{ec-p-2}) is satisfied for
\beq
|\alpha |\in (g_-,g_+).
\label{alfa-local-2}
\eeq

\begin{figure}[h!]
\centering
\includegraphics[width=8cm]{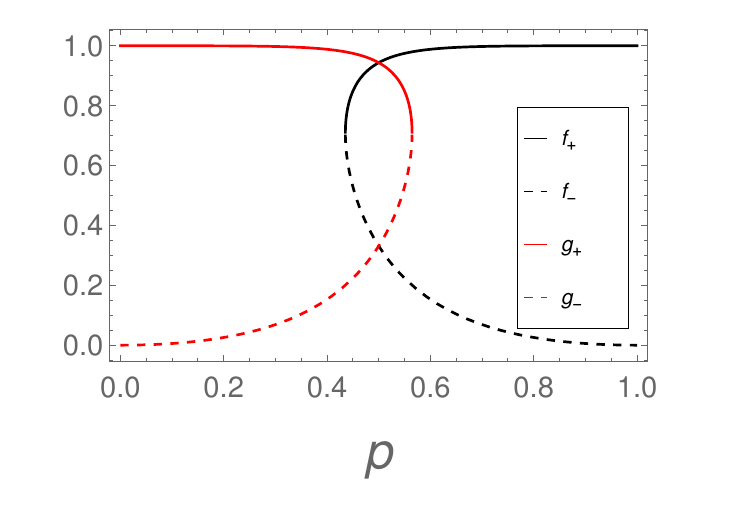}
\vspace{-0.5cm}
\caption{The range of $|\alpha |$ given by Eq. (\ref{alfa-local-1}) for which the state $ \rho^{a_1b_1}$ is inseparable is situated between the black curves. The range of $|\alpha |$ given by Eq. (\ref{alfa-local-2}) for which the state $ \rho^{a_2b_2}$ is inseparable is situated between the red curves. The region obtained by the intersection represents the range of both $|\alpha |$ and $p$ such that the two states are simultaneously inseparable.}
\label{reg-loc}
\end{figure}

We obtain the conditions for both states $ \rho^{a_1b_1}$ and $ \rho^{a_2b_2}$ to be inseparable (see Fig. \ref{reg-loc}):
\begin{itemize}
\item parameter $p$ satisfies $p\in (p_1,p_2)$, where $p_1$ and $p_2$ are given by Eqs. (\ref{expr-p1}) and (\ref{expr-p2});
\item $|\alpha |\in (0.331, 0.943)$;
\item the region obtained by the intersection of the black curves with the red ones in Fig. \ref{reg-loc} represents the range of both $|\alpha |$ and $p$ such that the two states are simultaneously inseparable.
\end{itemize}

It is easy to evaluate the concurrence of the initial state (\ref{entini}):
\beq
C(\ket{\psi}) = 2\, |\alpha |\, |\beta |.
\eeq

Further, we want to calculate $C(\ket{\psi})-[C(\rho^{a_1b_1})+C(\rho^{a_2b_2})]$ in the case when both states $ \rho^{a_1b_1}$ and $ \rho^{a_2b_2}$ are characterized by positive concurrence:
\beqa
&&C(\ket{\psi})-[C(\rho^{a_1b_1})+C(\rho^{a_2b_2})]=2|\alpha |\, |\beta |\, \frac{(1-p)^2p^2}{(1-p+p^2)^2}\nonumber \\
&&+ \frac{(1-p)^2+p^2+2p^2(1-p)^2}{(1-p+p^2)^2}>0.
\eeqa

This means that the concurrence of the initial pure state is greater than the sum of the concurrences of the two states generated by the local asymmetric broadcasting of entanglement:
\beq
C(\ket{\psi})>C(\rho^{a_1b_1})+C(\rho^{a_2b_2}).
\eeq

\section{Broadcasting of entanglement using nonlocal optimal universal asymmetric cloning machines}
\subsection{The output states generated by the nonlocal asymmetric broadcasting of entanglement}

We use the nonlocal asymmetric optimal cloning machine for dimension $d=4$ on the initial entangled state  \cite{Ghiu-pra-2003}:
\beq
 \ket{\psi} =\alpha \ket{00}+\beta \ket{11}
\eeq
in order to copy the two-qubit state at once.
The two asymmetric clones are described by the density operators \cite{Ghiu-pra-2003}:
\beq
\rho^{a_1b_1}=\frac{1}{1+3(p^2+q^2)}\left[ \left( 1-q^2+3p^2\right) \proj{\psi}{\psi}+q^2I\right]
\label{st-a1b1-neloc}
\eeq
and
\beq
\rho^{a_2b_2}=\frac{1}{1+3(p^2+q^2)}\left[ \left( 1-p^2+3q^2\right) \proj{\psi}{\psi}+p^2I\right].
\label{st-a2b2-neloc}
\eeq
By using Eq. (\ref{x-st}), we find that the two output states $\rho^{a_1b_1}$ and $\rho^{a_2b_2}$ are $X$ states.

\subsection{The concurrence of the output states generated by the nonlocal asymmetric broadcasting of entanglement}

We compute the concurrence of the output states with the help of formula (\ref{conc-X}):
\beqa
C(\rho^{a_1b_1})&=&2\max \bigg\{ 0, \frac{(1-q^2+3p^2)|\alpha |\, |\beta |-q^2}{1+3\, (p^2+q^2)}  \bigg\} \label{conc-neloc-a1b1}\\
C(\rho^{a_2b_2})&=&2\max \bigg\{ 0, \frac{(1-p^2+3q^2)|\alpha | \, |\beta |-p^2}{1+3\, (p^2+q^2)} \bigg\}.\label{conc-neloc-a2b2}
\eeqa

\begin{figure}[h!]
\centering
\includegraphics[width=6cm]{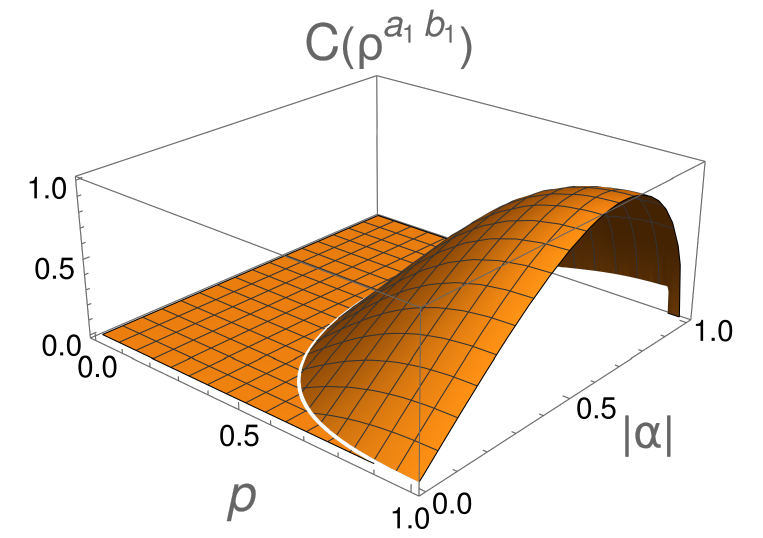}
\includegraphics[width=6cm]{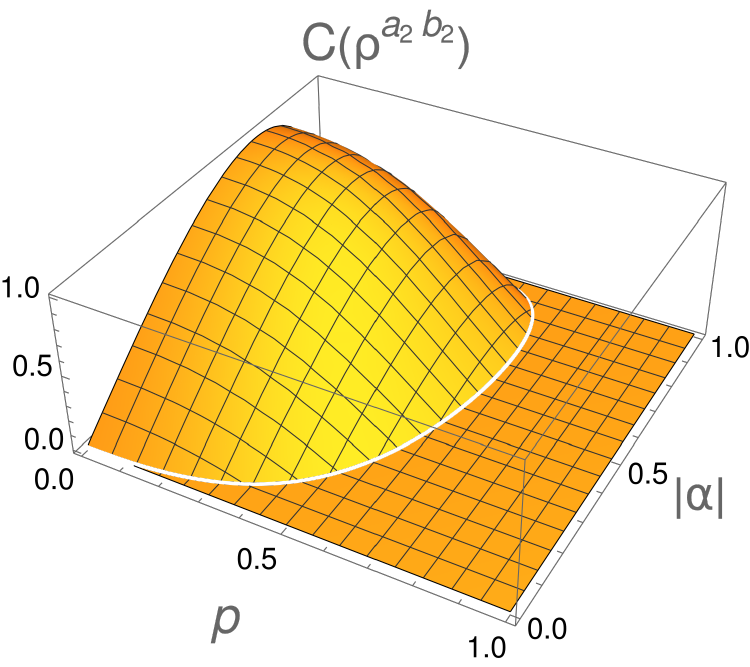}
\caption{Concurrence of $ \rho^{a_1b_1}$ (left) and $ \rho^{a_2b_2}$ (right) using a nonlocal asymmetric cloning machine.}
\label{fig-conc-neloc-a1-b1}
\end{figure}

In Fig. \ref{fig-conc-neloc-a1-b1} we plot the concurrence of the states $\rho^{a_1b_1}$ and $\rho^{a_2b_2}$, respectively.

Let us investigate when the maximum concurrence of the two states is obtained. This means that we need to evaluate the partial derivative of the concurrence with respect to $|\alpha|$:
\beqa
\frac{\partial C(\rho^{{a_1}{b_1}})}{\partial |\alpha|}&=&-\frac{2(-1+2 |\alpha|^2) p(1+p)}{\sqrt{1-|\alpha|^2}(2-3 p+3 p^2)} ; \nonumber \\
\frac{\partial C(\rho^{{a_2}{b_2}})}{\partial |\alpha|}&=&-\frac{2(-1+2 |\alpha|^2)(2-3 p+p^2)}{\sqrt{1-|\alpha|^2}(2-3 p+3 p^2)} .
\eeqa
In both cases we have that the maximum of the concurrence is obtained when the initial state is maximally entangled: $|\alpha|=\frac{1}{\sqrt{2}}$.

\begin{figure}[h!]
\centering
\includegraphics[width=6cm]{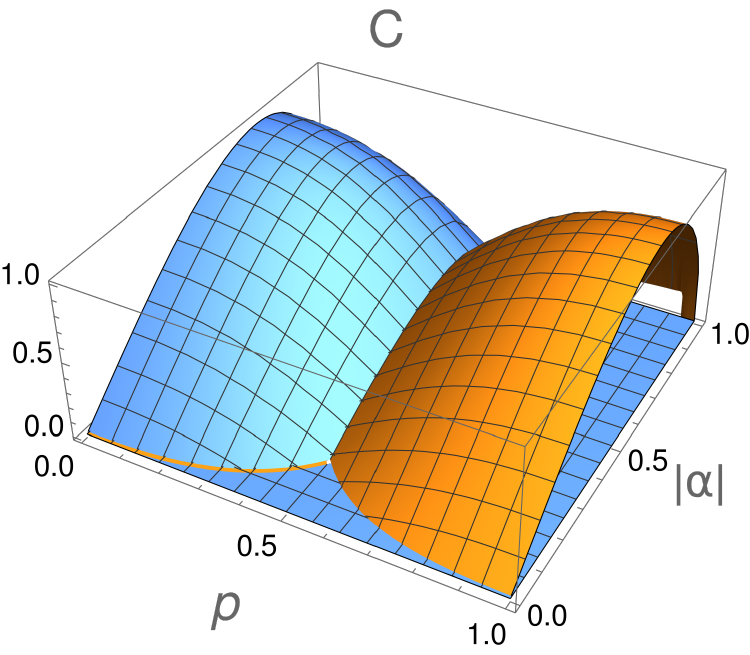}
\caption{A comparison between the concurrence of $ \rho^{a_1b_1}$ and the concurrence of $ \rho^{a_2b_2}$ when a nonlocal asymmetric cloning machine is employed.}
\label{fig-conc-neloc-a1b1-a2b2}
\end{figure}

We analyze now in which conditions the two output states are simultaneously inseparable: see Fig. \ref{fig-conc-neloc-a1b1-a2b2}.
By using Eq. (\ref{conc-neloc-a1b1}), the condition $C(\rho^{a_1b_1})> 0 $ is equivalent to
\beq
|\alpha |^4-|\alpha |^2+\frac{q^4}{(1-q^2+3p^2)^2}< 0.
\label{in-neloc-1}
\eeq
This inequality is verified only if $p\in (\frac{1}{3},1]$. With the notation
\beq
\xi_\pm =\sqrt{\frac{1}{2}\left[ 1\pm \sqrt{1-\frac{4q^4}{(1-q^2+3p^2)^2}}\right]},
\eeq
we obtain that Eq. (\ref{in-neloc-1}) leads to
\beq
|\alpha |\in (\xi_-,\xi_+).
\label{alfa-nelocal-1}
\eeq

The condition $C(\rho^{a_2b_2})> 0 $ can be rewritten with the help of Eq. (\ref{conc-neloc-a2b2}) as follows:
\beq
|\alpha |^4-|\alpha |^2+\frac{p^4}{(1-p^2+3q^2)^2}< 0.
\label{in-neloc-2}
\eeq
This leads to the condition $p\in [0, \frac{2}{3}) $. If we denote
\beq
\eta_\pm =\sqrt{\frac{1}{2}\left[ 1\pm \sqrt{1-\frac{4p^4}{(1-p^2+3q^2)^2}}\right]},
\eeq
then one obtains
\beq
|\alpha |\in (\eta_-,\eta_+).
\label{alfa-nelocal-2}
\eeq

\begin{figure}[h!]
\centering
\includegraphics[width=8cm]{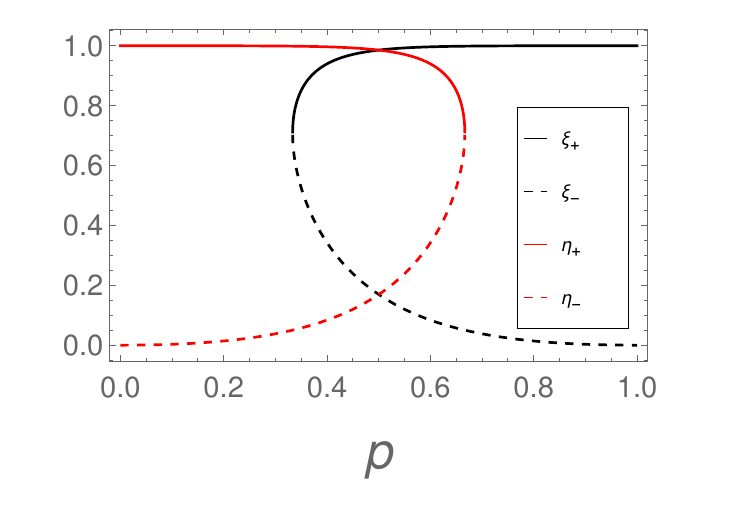}
\vspace{-0.5cm}
\caption{The range of $|\alpha |$ given by Eq. (\ref{alfa-nelocal-1}) for which the state $ \rho^{a_1b_1}$ is inseparable is situated between the black curves. The range of $|\alpha |$ given by Eq. (\ref{alfa-nelocal-2}) for which the state $ \rho^{a_2b_2}$ is inseparable is situated between the red curves. The region obtained by the intersection represents the range of both $|\alpha |$ and $p$ such that the two states are simultaneously inseparable.}
\label{reg-neloc}
\end{figure}

We obtain the conditions for both states $ \rho^{a_1b_1}$ and $ \rho^{a_2b_2}$ to be inseparable (see Fig. \ref{reg-neloc}):
\begin{itemize}
\item parameter $p$ satisfies $p\in (\frac{1}{3},\frac{2}{3})$;
\item $|\alpha |\in (0.169, 0.985)$;
\item the region obtained by the intersection of the black curves with the red ones in Fig. \ref{reg-neloc} represents the range of both $|\alpha |$ and $p$ such that the two states are simultaneously inseparable.
\end{itemize}

 Let us evaluate now $C(\ket{\psi})-[C(\rho^{a_1b_1})+C(\rho^{a_2b_2})]$ in the case when both states $ \rho^{a_1b_1}$ and $ \rho^{a_2b_2}$ are inseparable:
\beq
C(\ket{\psi})-[C(\rho^{a_1b_1})+C(\rho^{a_2b_2})]=\frac{1-2(1+|\alpha |\, |\beta |)p(1-p)}{2-3p+p^2}.
\label{difconc-neloc}
\eeq
Due to the fact that the function (\ref{difconc-neloc}) has a minimum for $|\alpha |=\frac{1}{\sqrt 2}$, we obtain:
\beq
C(\ket{\psi})-[C(\rho^{a_1b_1})+C(\rho^{a_2b_2})]\ge \frac{1-3p+3p^2}{2-3p+p^2}>0.
\eeq

In conclusion, we have found that the concurrence of the initial pure state is greater than the sum of the concurrences of the two states generated by the nonlocal asymmetric broadcasting of entanglement:
\beq
C(\ket{\psi})>C(\rho^{a_1b_1})+C(\rho^{a_2b_2}).
\eeq

\section{A comparison between asymmetric broadcasting of entanglement using local and nonlocal cloning machines}

If one compares the range of the parameters $\alpha $ and $p$ in the case of local broadcasting of entanglement obtained in Section 3.2, for which the two output states are entangled, with the range of $\alpha $ and $p$ obtained in Section 4.2 (nonlocal broadcasting), we arrive at the conclusion that the nonlocal broadcasting of entanglement generated wider intervals for these parameters compared with the local case. This leads to two advantages of nonlocal broadcasting over local broadcasting:
\begin{itemize}
\item a wider class of initial states (\ref{entini}) can be broadcast to generate two inseparable output states;
\item a larger class of optimal universal asymmetric cloning machines can be employed (larger $p$).
\end{itemize}

On the other hand, we want to compare of the concurrences of the state $\rho^{a_1b_1}$ when local cloning machines are applied and when nonlocal cloning machines are used:
\beqa
C(\rho^{a_1b_1}_{\mbox{nonloc}})-C(\rho^{a_1b_1}_{\mbox{loc}})&=&2\, \frac{(1-q^2+3p^2)|\alpha |\, |\beta |-q^2}{1+3\, (p^2+q^2)}-2\, \frac{4p^2|\alpha |\, |\beta |-q^2(1+p^2)}{(1+p^2+q^2)^2}\nonumber \\
&=&\frac{p(1-p)^2(1-p+p^2+p^3)}{2\, (1-p+p^2)^2(2-3p+3p^2)}\, (1+4\, |\alpha|\, |\beta |).\label{dif-C-1}
\eeqa
We can rewrite
\beq
2-3p+3p^2=(1-p)^2+1-p+2p^2>0,
\eeq
which together with Eq. (\ref{dif-C-1}) leads to
\beq
C(\rho^{a_1b_1}_{\mbox{nonloc}})>C(\rho^{a_1b_1}_{\mbox{loc}})
\label{comp-conc-a1b1}
\eeq
for any $\alpha $ and $p$.

On the other hand, we have
\beqa
C(\rho^{a_2b_2}_{\mbox{nonloc}})-C(\rho^{a_2b_2}_{\mbox{loc}})&=&2\, \frac{(1-p^2+3q^2)|\alpha |\, |\beta |-p^2}{1+3\, (p^2+q^2)}-2\, \frac{4q^2|\alpha |\, |\beta |-p^2(1+q^2)}{(1+p^2+q^2)^2}\nonumber \\
&=&\frac{q(1-q)^2(1-q+q^2+q^3)}{2\, (1-q+q^2)^2(2-3q+3q^2)}\, (1+4\, |\alpha|\, |\beta |).\label{dif-C-2}
\eeqa
From Eq. (\ref{dif-C-2}) and by writing $ 2-3q+3q^2= (1-q)^2+1-q+2q^2>0 $, we find that
\beq
C(\rho^{a_2b_2}_{\mbox{nonloc}})>C(\rho^{a_2b_2}_{\mbox{loc}})
\label{comp-conc-a2b2}
\eeq
for any $\alpha $ and $p$.

This means that the concurrence is greater in the case of nonlocal asymmetric broadcasting than in the case of local asymmetric broadcasting.

\section{Application of the asymmetric broadcasting of entanglement: the use of the output states in quantum teleportation }

\subsection{The use of the output states of the local broadcasting of entanglement as quantum channels in teleportation}

 In this section, we investigate the possiblity of using the output states of the local broadcasting of entanglement, i.e. the states $\rho^{a_1b_1}$ of Eq. (\ref{st-a1b1-loc}) and $\rho^{a_2b_2}$ of Eq. (\ref{st-a2b2-loc}),  as quantum channels in teleportation. We consider only the case when the states  $\rho^{a_1b_1}$  and $\rho^{a_2b_2}$ are inseparable.
 Both of these states are $X$ states characterized by $\rho_{23} =0$ and $\rho_{22}=\rho_{33}$.

By using  Eq. (\ref{st-a1b1-loc}), we have for the state $\rho^{a_1b_1}$
\beq
\rho^{a_1b_1}_{22}-\frac{1}{4}= -\frac{p^2}{(1+p^2+q^2)^2}<0.
\eeq

 Since the state is inseparable, $\rho_{23} =0$, $\rho_{22}=\rho_{33}$ and $\rho_{22}<\frac{1}{4}$, it means that all the conditions required in the hypothesis of the Theorem given in Section 2, are satisfied for the state $\rho^{a_1b_1}$.

With the help of Eq. (\ref{st-a2b2-loc}) we obtain for the state $\rho^{a_2b_2}$
\beq
\rho^{a_2b_2}_{22}-\frac{1}{4}= -\frac{q^2}{(1+p^2+q^2)^2}<0,
\eeq
i.e., we can again apply the Theorem from Section 2.

This leads to the conclusion that all the inseparable states $\rho^{a_1b_1}$ of Eq. (\ref{st-a1b1-loc}) and $\rho^{a_2b_2}$ of Eq. (\ref{st-a2b2-loc}) are useful for teleportation.

\begin{figure}[h!]
\centering
\includegraphics[width=6cm]{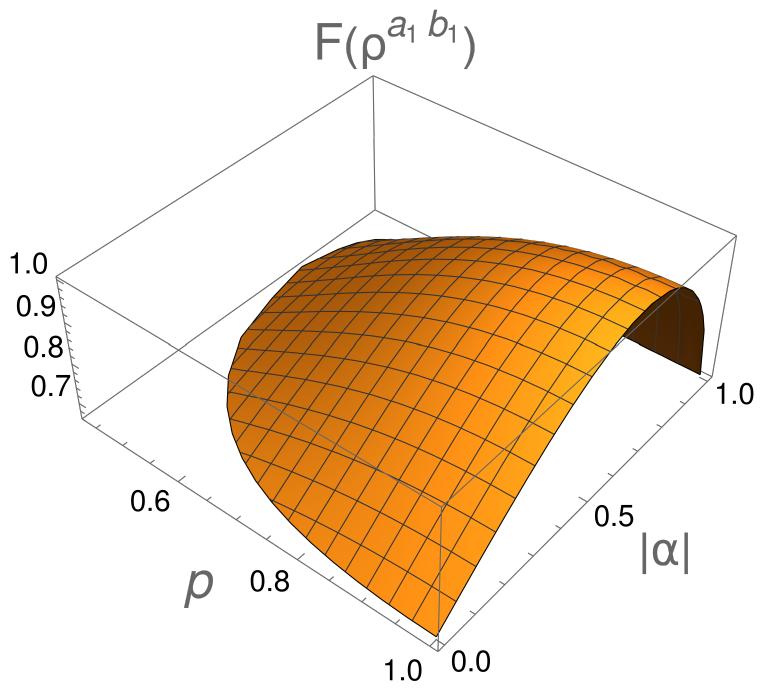}
\includegraphics[width=6cm]{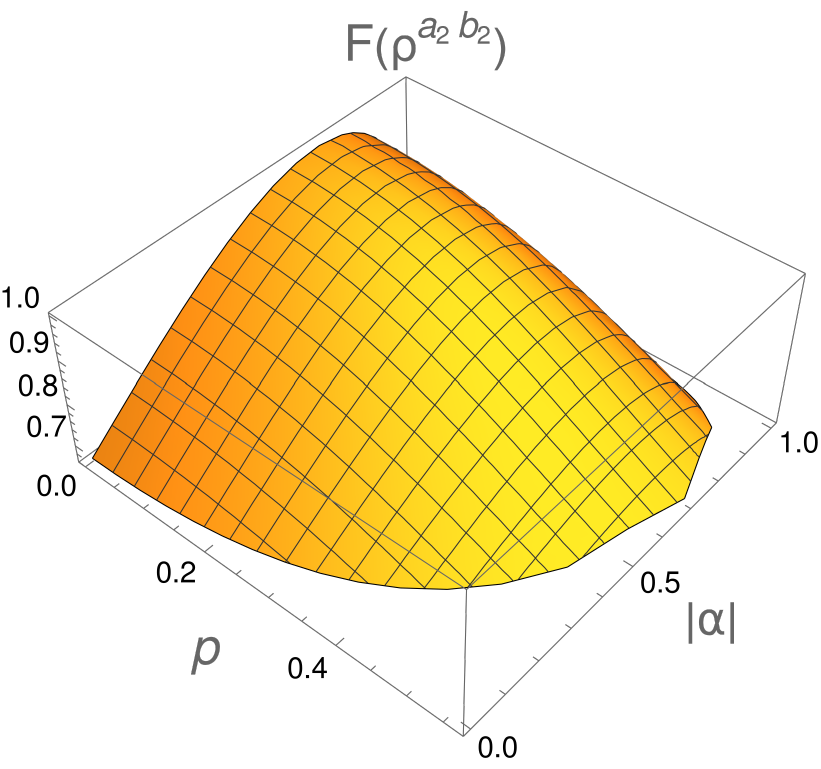}
\caption{Fidelity of teleportation, when the states $\rho^{a_1b_1}$ and $\rho^{a_2b_2}$, given by the local broadcasting of entanglement, are used as quantum channels.}
\label{fig-fid-loc}
\end{figure}

Having obtained the concurrence of the states $\rho^{a_1b_1}$ and $\rho^{a_2b_2}$ in Section 3.2, we are able to evaluate the fidelity of teleportation with the help of Eq. (\ref{leg-fid-concurr}). In Fig. \ref{fig-fid-loc}, we plot the fidelity of quantum teleportation, when the states $\rho^{a_1b_1}$ and $\rho^{a_2b_2}$ are used as quantum channels.

\subsection{The use of the output states of the nonlocal broadcasting of entanglement as quantum channels in teleportation }

We investigate the case when  the entangled two-qubit states $\rho^{a_1b_1}$ of Eq. (\ref{st-a1b1-neloc}) and $\rho^{a_2b_2}$ of Eq. (\ref{st-a2b2-neloc}), generated in the process of nonlocal broadcasting of entanglement, are used as quantum channels in the standard teleportation. These two mixed states are $X$ states, being characterized by the matrix elements $\rho_{23} =0$ and $\rho_{22}=\rho_{33}$.
In addition, we have
\beqa
\rho^{a_1b_1}_{22}-\frac{1}{4}&=&-\frac{1+p^2+(1+p)^2}{1+3(p^2+q^2)}<0; \nonumber \\
\rho^{a_2b_2}_{22}-\frac{1}{4}&=&-\frac{1+q^2+(1+q)^2}{1+3(p^2+q^2)}<0.
\eeqa

By applying the Theorem from Section 2, we arrive at the conclusion that all the inseparable states $\rho^{a_1b_1}$ of Eq. (\ref{st-a1b1-neloc}) and $\rho^{a_2b_2}$ of Eq. (\ref{st-a2b2-neloc}) can be used as quantum channels for the standard teleportation.

\begin{figure}[h!]
\centering
\includegraphics[width=6cm]{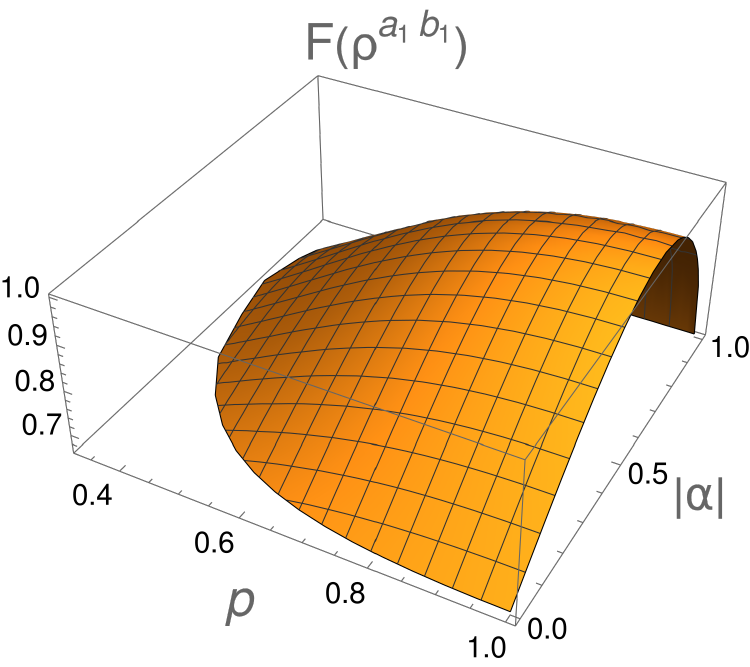}
\includegraphics[width=6cm]{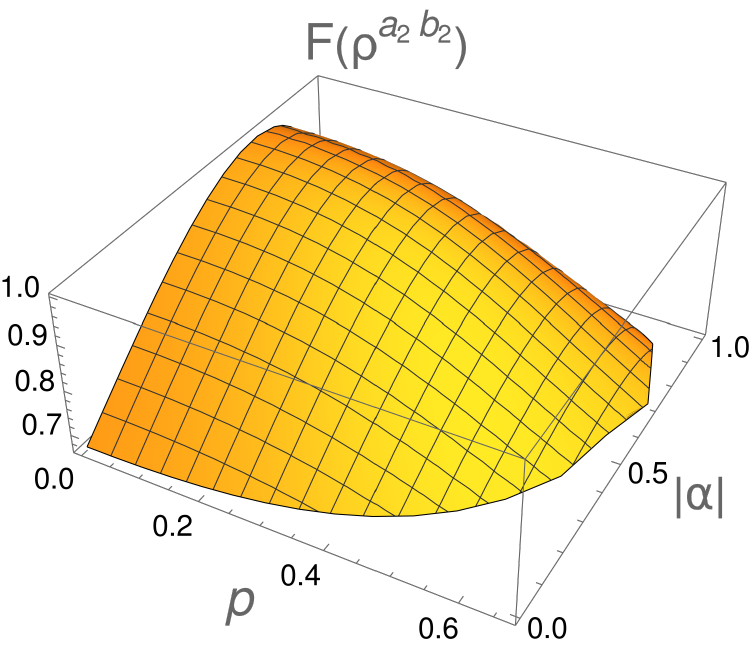}
\caption{Fidelity of teleportation, when the states $\rho^{a_1b_1}$ and $\rho^{a_2b_2}$, given by the nonlocal broadcasting of entanglement, are used as quantum channels.}
\label{fig-fid-neloc}
\end{figure}

Since the concurrence of the states $\rho^{a_1b_1}$ and $\rho^{a_2b_2}$ was determined in Sec. 4.2, we can obtain the expression of the fidelity of teleportation with the help of Eq. (\ref{leg-fid-concurr}). The dependence of the fidelity of teleportation on $|\alpha |$ and $p$, obtained when states $\rho^{a_1b_1}$ of Eq. (\ref{st-a1b1-neloc}) and $\rho^{a_2b_2}$ of Eq. (\ref{st-a2b2-neloc}) are used as quantum channels, is shown in Fig. \ref{fig-fid-neloc}.

\section{A comparison between the fidelity of teleportation when the quantum channels are obtained by asymmetric local and nonlocal broadcasting of entanglement}

In Sec. 5, we demonstrated that the concurrence of the state $\rho^{a_1b_1}$ is greater when nonlocal broadcasting of entanglement is used than when local broadcasting is used (see Eq. (\ref{comp-conc-a1b1})). The same conclusion is valid for the state $\rho^{a_2b_2}$, as one can see from Eq. (\ref{comp-conc-a2b2}):
\beqa
C(\rho^{a_1b_1}_{\mbox{nonloc}})&>&C(\rho^{a_1b_1}_{\mbox{loc}}); \nonumber \\
C(\rho^{a_2b_2}_{\mbox{nonloc}})&>&C(\rho^{a_2b_2}_{\mbox{loc}}). \label{comp-c-gen}
\eeqa

The two inseparable states $\rho^{a_1b_1}$ and $\rho^{a_2b_2}$ that are used as quantum channels in teleportation are $X$ states, being characterized by $\rho_{23}=0$ and $\rho_{22}=\rho_{33}< \frac{1}{4}$. This means that we can apply the Theorem from Section 2, namely Eq. (\ref{leg-fid-concurr}), which gives the relation between the maximal fidelity of teleportation and the concurrence, and this leads to:
\beqa
&&F_{\mbox{max}}(\rho^{a_1b_1}_{\mbox{nonloc}})>F_{\mbox{max}}(\rho^{a_1b_1}_{\mbox{loc}})\nonumber \\
&&F_{\mbox{max}}(\rho^{a_2b_2}_{\mbox{nonloc}})>F_{\mbox{max}}(\rho^{a_2b_2}_{\mbox{loc}}),\nonumber
\eeqa
where we have used the two inequalities (\ref{comp-c-gen}).

We arrive at the conclusion that the maximal fidelity of teleportation when the quantum channels are obtained by broadcasting of entanglement using nonlocal cloning machines is always greater than  when local cloning machines are employed.

\section{Conclusions}

The first significant finding is a theorem establishing a relationship between the concurrence of a special $X$ state and the maximal fidelity of teleportation obtained when this state is used as a quantum channel. Furthermore, by applying the local asymmetric broadcasting of entanglement, we generate two output states, which are studied in detail by evaluating the concurrences and  finding the conditions for the two states to be simultaneously inseparable. Further, the same procedure is employed for the nonlocal asymmetric broadcasting of entanglement.
We have proved that the range of the parameters $\alpha $ (which describes the initial entangled state) and $p$ (which characterizes the cloning machines) in the case of nonlocal broadcasting of entanglement is larger than the range of the two parameters when local asymmetric broadcasting of entanglement is performed, if we ask for the two output states to be entangled. In addition, we have found that the concurrence is greater in the case of nonlocal broadcasting.

By applying the theorem proved in this paper, we show that all the inseparable output states of the local broadcasting of entanglement can be used as quantum channels in teleportation.  A similar analysis is presented for the case when the two output states are obtained in the nonlocal broadcasting of entanglement. Finally, we make a comparison  between the fidelities of teleportation when the quantum channels are obtained by asymmetric broadcasting of entanglement using local and nonlocal cloning machines, and show that better results are obtained in the case of nonlocal broadcasting of entanglement.

\section*{Acknowledgements}
This work was supported by a grant of the Romanian Ministry of Research, Innovation and Digitalization, CNCS-UEFISCDI, project number PN-III-P4-ID-PCE-2020-1142, within PNCDI III.

\end{document}